# Synthesis of strain-relaxed Ge-Sn alloys using ion implantation and pulsed laser melting

Tuan T. Tran [1*], Quentin Hudspeth [2], Yining Liu [3], Lachlan A. Smillie [1], Buguo Wang [4], Renaud A. Bruce [5,6], Jay Mathews [3], Jeffrey M. Warrender [2], J. S. Williams [1]

[1] *Department of Electronic Materials Engineering, Research School of Physics and Engineering, Australian National University, Canberra, ACT 0200, Australia.*

[2] *U.S. Army Combat Capabilities Development Command – Armament Center Watervliet, New York 12189, USA.*

[3] *Department of Electro-Optics and Photonics, University of Dayton, Dayton, OH 45469, USA.*

[4] *Department of Engineering Physics, Air Force Institute of Technology, Wright-Patterson AFB, OH 45433, USA.*

[5] *Department of Physics, Morehouse College, Atlanta, Georgia 30314, USA.*

[6] *Department of Physics, University of Dayton, Dayton, OH 45469, USA.*

(*) corresponding author: tuan.tran@physics.uu.se

Ge-Sn alloys with a sufficiently high concentration of Sn is a direct bandgap group IV material. Recently, ion implantation followed by pulsed laser melting has been shown to be a promising method to realize this material due to its high reproducibility and precursor-free process. A Ge-Sn alloy with ∼9 at. % Sn was shown to be feasible by this technique. However, the compressive strain, inherently occurring in heterogeneous epitaxy of the film, evidently delays the material from the direct bandgap transition. In this report, an attempt to synthesise a highly-relaxed Ge-Sn alloy will be presented. The idea is to produce a significantly thicker film with a higher implant energy and doses. X-ray reciprocal space mapping confirms that the material is largely-relaxed. The peak Sn concentration of the highest dose sample is 6 at. % as determined by Rutherford backscattering spectrometry. Cross-sectional transmission electron microscopy shows unconventional defects in the film as the mechanism for the strain relaxation. Finally, a photoluminescence (PL) study of the strain-relaxed alloys shows photon emission at a wavelength of $2045\,nm$, suggesting an active incorporation of Sn concentration of 6 at. %. The results of this study pave the way to producing high quality relaxed GeSn alloy using an industrially scalable method.

## Introduction

Group IV semiconductor alloys have been a research focus for many years due to their potential to improve the performance of electronic devices. For example, silicon-germanium alloys (Si-Ge) can be used as active channel materials in metal-oxide-semiconductor field effect transistors (MOSFETs) due to the improved carrier mobility [1], or as the source/drain stressors in p-MOSFETs [2]. Recently, considerable attention has been directed to another group IV alloy, namely germanium-tin (Ge-Sn), because this alloy is a direct bandgap semiconductor at a certain composition of Sn.

The material has been demonstrated using various non-equilibrium techniques, such as molecular beam epitaxy [3-5], solid phase epitaxy [6], sputter deposition [7], and chemical vapour deposition (CVD) [8-10]. For example, after developing a stable Sn precursor (phenyl-Sn-deuterium), the group of Kouvetakis and Menendez successfully synthesised the Ge-Sn alloy using CVD [11]. The first direct bandgap photoluminescence at room temperature from a Ge-Sn alloy was reported by the same group about a decade later in 2010 [12]. Another precursor option is tin chloride ($SnCl_4$), which is a stable liquid and can be introduced into the reaction chamber by a vapour station [13]. With both of these precursors, light amplification and stimulated emission from the resulting materials have been demonstrated [8,14].

Compressive strain is an important issue in realizing the direct bandgap Ge-Sn alloy. Due to a large lattice mismatch between α-Sn (6.489 Å), Ge (5.646 Å) and Si (5.431 Å) [15], the Ge-Sn alloy grown pseudo-morphicallly on Si or Ge substrates often endures a great compressive strain that limits the soluble Sn concentration, reduces crystal quality and delays the transition towards a direct bandgap material. Theoretical studies show that a significantly higher Sn concentration is necessary for a partially strained alloy, whereas, the material does not have a direct bandgap at any composition for a fully strained material [16]. CVD growth of the strained-relaxed Ge-Sn alloy on Si has also been demonstrated through the introduction of Lomer dislocations locally at the GeSn/Si interface [8,17-19] or the growth of Ge-Sn nanostructures, such as nanowires [20].

Ion implantation followed by pulsed laser melting and solidification (PLM), i.e. ion beam synthesis, is a promising approach because the process is highly controllable, precursor-free, and hence virtually free of contamination. Using ion beam synthesis, we are able to produce a high quality Ge-Sn crystal with ~9 at.% Sn [21]. This was done by using liquid nitrogen implantation [22] and a capping layer of SiO2 to overcome the porosity issues of the implanted Ge [23]. The film was shown to have the thickness of $\sim 70\ nm$ and pseudo-morphically grown on the Ge substrate. Reciprocal space mapping showed the material was fully-strained with lattice expansion normal to the surface. As an attempt to synthesise a strain-relaxed Ge-Sn alloy, in this study the material is produced at a higher implant energy for a thicker film. Physical properties of the films are characterised by Rutherford backscattering spectrometry (RBS), X-ray diffraction (XRD), transmission electron microscopy (TEM) and finally photoluminescence spectroscopy (PL).

**Experiment**

An optimal and laterally uniform laser fluence is important to achieve high quality material. It must be high enough to melt through the disordered/amorphous layer so that the resolidified layer has a crystalline seed from which to grow epitaxially. On the other hand, excessive laser fluence will induce longer melt duration, causing bulk and surface segregation of the implanted impurities, and possibly even surface ablation. Therefore, simulation and preliminary experiments are necessary to find the optimal laser fluence for each thickness of the amorphous Ge-Sn layer. A finite-difference code that performs one-dimensional heat flow calculations including phase changes was used to calculate the predicted melt depth as a function of the laser fluence. These predicted laser fluences were then used as reference for the experimental PLM step. For this purpose, bulk Ge (100) substrates were implanted with Sn ions at a fixed dose of $1 \cdot 10^{16}\ cm^{-2}$ and at an increasing energy of 120 keV, 250 keV, 350 keV and 450 keV. All implants were done at the substrate temperature of ~77 K to suppress the formation of porosity in Ge. The thickness of the as-implanted Ge-Sn layer was 130 nm, 240 nm, 300 nm and 370 nm, respectively, as determined by RBS. Each sample received one shot from a frequency-tripled Nd:YAG laser (Ekspla), 355 nm, having a FWHM pulse duration of 6 ns. The beam was spatially homogenized, and passed through a 4 mm × 4 mm aperture. The melt duration was monitored with time-resolved reflectivity using an Ar⁺ laser, and fluences were calibrated by measuring melt durations on as-received Si and Ge. Based on the predicted melt depths obtained from calculations, a series of systematically increased laser fluences was applied to each sample of different implant energy as follows: $0.32 - 0.41\ J \cdot cm^{-2}$ for the 120 keV sample, $0.41 - 0.55\ J \cdot cm^{-2}$ for the 250 keV sample, $0.49 - 0.65\ J \cdot cm^{-2}$ for the 350 keV sample and $0.62 - 0.78\ J \cdot cm^{-2}$ for the 450 keV sample. RBS/C was again used to study the crystal quality of the PLM samples and to identify the proper laser fluence for each thickness of the Ge-Sn layer.

For detailed study of the Ge-Sn alloy, an implant energy of 350 keV was selected for a series of implant doses of $5.0 - 6.8 \cdot 10^{16}$ cm$^{-2}$. The substrates used at this stage were relaxed Ge-on-Si. Prior to the implantation, a ~40 nm capping layer of SiO$_2$ was deposited on all substrates by plasma-enhanced CVD to prevent Ge from becoming porous [23]. After implantation, the samples were dipped into buffered hydrofluoric acid (HF:NH$_4$F = 1:7) for ~30 sec to remove the residual SiO$_2$ capping layer. Because the intermixed oxygen from the capping layer after implantation can considerably hinder good quality regrowth [21], ~30 nm of the sample surface was removed by plasma etching (RIE) to minimise the undesired oxygen before PLM.

The laser fluence of $0.52 - 0.62$ J·cm$^{-2}$ was used for the 350 keV implant as determined from the preliminary experiments. A 2 MeV He$^+$ beam from the RBS system was used to characterize the samples both after implantation and after PLM. To study the strain in the Ge-Sn samples, a high resolution PANalytical X'Pert XRD system was used for reciprocal space mapping by undertaking a series of asymmetric $\omega/2\theta$ scans with slightly different $\omega$ offsets along the $[\bar{2}\bar{2}4]$ crystal axis. Bright field and high resolution transmission electron microscopy (HR-TEM) was done with a JEOL JEM-2100F system. Finally, a photoluminescence (PL) study of the Ge-Sn alloy was done with a $976\ nm$ diode laser as a pump source at the excitation power of $1 - 15$ W. The laser was chopped at a frequency of ~300 Hz, and the modulated PL signal was focused into a Horiba MicroHR spectrometer ($f = 140$ mm) with a grating of 300 lines/mm and 2 µm blaze wavelength. A LN$_2$-cooled extended-InGaAs single channel detector ($1.3 - 2.4$ µm) was used for detection and connected to a Stanford Research SR830 lock-in amplifier. This range of detection is necessary because the indirect bandgap of the pristine Ge material is ~0.66 eV (equivalent to an infrared wavelength of 1880 nm). At the transition to direct bandgap, a bandgap energy of $< 0.61$ eV is expected. During the measurement, samples were kept at room temperature under atmospheric ambient.

**Results and discussion**

Fig. 1(a) shows the predicted melt depth as a function of laser fluence from heat flow calculations for 4 different amorphous Ge thicknesses on a crystalline substrate: 130 nm (120 keV Sn implant), 240 nm (250 keV implant), 300 nm (350 keV implant) and 370 nm (450 keV implant).

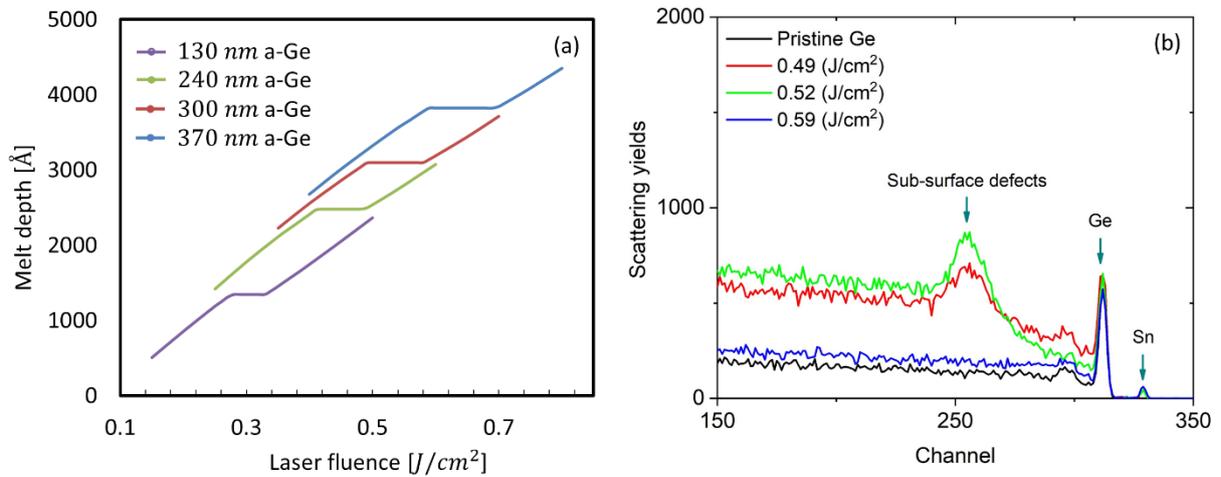

Fig. 1: LIMP simulation of the melt depth as a function of the laser fluence for the amorphous thickness of 130 nm, 240 nm, 300 nm and 370 nm (a). RBS/C spectra of the PLM spots on the 350 keV implant sample at different laser fluence (b).

The RBS/channelled (RBS/C) spectra for the laser spots of different fluences are shown in Fig. 1(b) for the 350 keV samples. Spectra of the 0.49 J/cm$^2$ and 0.52 J/cm$^2$ samples show significantly higher dechanneling than the others, corresponding to a higher degree of disorder. Such a signature is indicative of incomplete melting of the amorphous or ion-damaged layer, which then resolidifies as a defective single crystal or a polycrystalline layer. Spectra for the higher fluences, such as 0.59 J/cm$^2$, do not show this behaviour, and hence we conclude that the laser fluence for the 350 keV Sn implant should be above 0.59 J/cm$^2$ to ensure melting through the entire amorphous and damaged layer. By the same procedure, the sufficient laser fluences for the 120 keV, 250 keV and 450 keV implants were identified to be 0.36, 0.46 and 0.59 J/cm$^2$, respectively. Beyond 0.7 J/cm$^2$ ablation appeared to initiate on the surface, which can remove material and generate defects in the PLM samples. According to our experimental data it is possible to melt the samples implanted at the energy of 450 keV without exceeding the ablation limit; however only samples implanted at 350 keV will be shown for the rest of this report.

Fig. 2 presents the RBS spectra of the $6 \cdot 10^{16}$ cm$^{-2}$ sample, including the as-implanted spectrum (black), the PLM/channelled (red) and the PLM/random (green), as well as the simulation using the SIMNRA code (magenta). The simulation curve (to best fit the as-implanted spectrum) shows that after implantation the total amorphous thickness is ~360 nm and the peak Sn concentration is ~8.5 at.%. After PLM, the RBS/random spectrum shows a small change in the Sn profile, with the Sn distribution spread more uniformly throughout the depth of the Ge-Sn layer. Using SIMNRA, the thickness of this uniform Ge-Sn layer is estimated to be ~110 nm and the Sn concentration is ~7.0 at.%. The RBS/channelled spectrum shows no evident Sn surface peak, indicating the absence of surface segregation. By using both the RBS/channelled and the RBS/random spectra of the PLM sample, the substitutional fraction of the implanted Sn ions is calculated to be ~75%. Hence, the substitutional Sn concentration of the $6 \cdot 10^{16}\ cm^{-2}$ sample is ~5.25 at.%. Using the same method, the substitutional Sn concentration of the $5 \cdot 10^{16}\ cm^{-2}$ and the $6.8 \cdot 10^{16}\ cm^{-2}$ samples are estimated to be ~4.8 at.% and ~6 at.%, respectively.

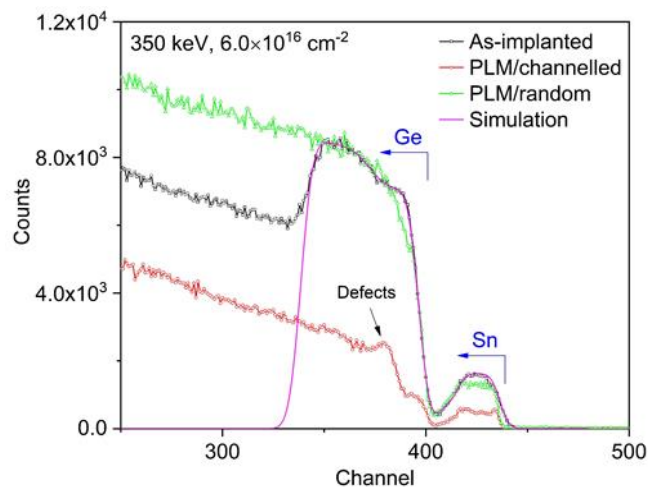

Fig. 2: RBS spectra in channelling (RBS/C) and random geometry (RBS/R) of the samples implanted at the energy of 350 keV and the dose of $6 \cdot 10^{16}\ cm^{-2}$.

In the RBS/channelled spectrum of Fig. 2, a noteworthy feature is the peak below the surface in the Ge part of the spectrum. While this peak does suggest a region of defects below the sample surface, it will be shown later that the subsurface defective regions in this sample are not associated with large amorphous blobs, such as the ones previously reported in Ref. [21] due to oxygen contamination. Rather, these are the first crystal defects to occur at the solid/liquid interface during the resolidification to accommodate (and relax) the excessive compressive stress built up in the layer. This process eventually leads to strain relaxation in the Ge-Sn alloy as will be shown in the following figure.

The XRD-$\omega/2\theta$ scan on the asymmetric $(\bar{2}\bar{2}4)$ planes of the $6 \cdot 10^{16}$ cm$^{-2}$ sample is shown in Fig. 3(a). The Ge substrate peak at 0 s is for reference. Due to the lattice expansion, a single XRD peak from the Ge-Sn layer is at a lower Bragg angle. Reciprocal space mapping using the same XRD system was employed to fully characterise the lattice parameters of the alloy, such as the in-plane and out-of-plane strain as shown in Fig. 3(b). The most noteworthy feature in this figure is that the Ge-Sn peak is located along a diagonal axis relative to the Ge peak of the substrates. The full relaxation axis is determined by way of the in-plane and out-of-plane lattice constants ($a$ and $c$, respectively) having the same values. The RSM result therefore indicates that the Ge-Sn layer is fully relaxed because its peak is situated on the axis of full relaxation.

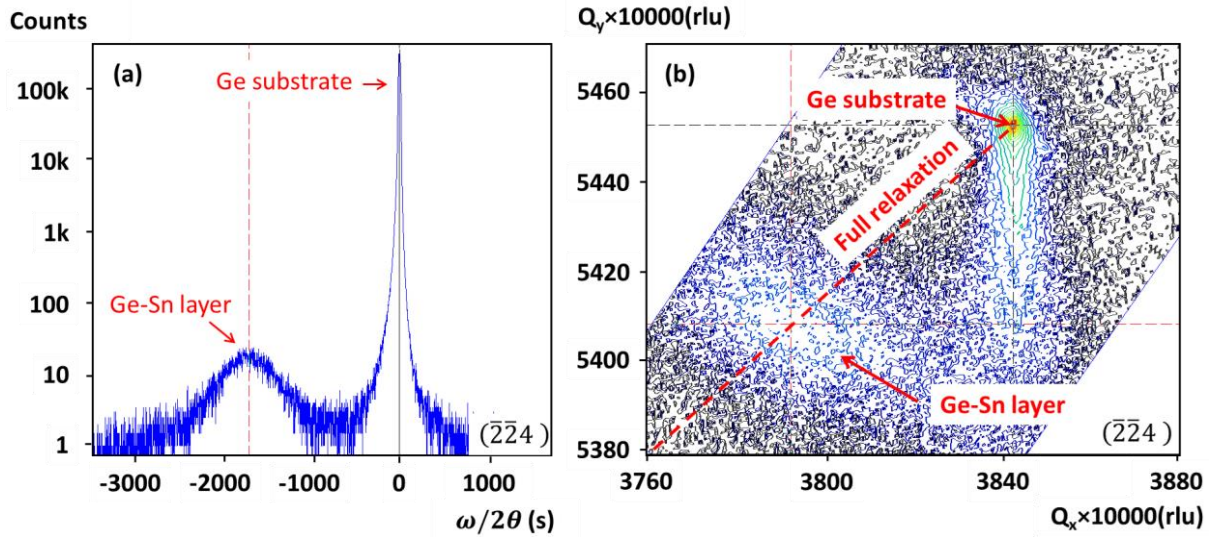

Fig. 3: XRD-$\omega/2\theta$ scan on the ($\bar{2}\bar{2}4$) plane (a) and reciprocal space mapping on the ($\bar{2}\bar{2}4$) planes of the $6.0 \cdot 10^{16}$ cm$^{-2}$ sample (b).

From the RSM measurement, lattice constants of the Ge and Ge-Sn crystal can be calculated with the following equation: $c = \frac{1}{Q_y(rlu)} \frac{\lambda}{2} l$, where $c$ is the lattice constant of a diamond cubic structure, $Q_y$ is the reciprocal constant taken from the RSM measurement, $\lambda = 1.54$ Å is the wavelength of the copper $K_\alpha$ emission from the X-ray source and $l = 4$ is the third Miller indexes of the ($\bar{2}\bar{2}4$) planes. By applying this equation, the lattice constant of the Ge substrate is ~5.65 Å, consistent with existing data in the literature [15,24], whereas the lattice constant of the Ge-Sn layer is ~5.691 Å for the $6 \cdot 10^{16}$ $cm^{-2}$ sample and ~5.695 Å for the $6.8 \cdot 10^{16}$ $cm^{-2}$ sample. The lattice constant is related to the concentration of the constituents as represented by Vegard's law: $a_{Ge-Sn} = x \cdot a_{Sn} + (1-x) \cdot a_{Ge} + x \cdot (1-x) \cdot b$, where $x$ is the Sn concentration and b is the bowing parameter, which is 0.047 Å according to Ref. [25]. Using the reference lattice constants of α-Sn and Ge and the lattice constant of the Ge-Sn alloys from the XRD measurements, the Sn concentration of the $6 \cdot 10^{16}$ $cm^{-2}$ and the $6.8 \cdot 10^{16}$ $cm^{-2}$ alloys is ~5.1 at.% and ~5.6 at.%, respectively. The substitutional Sn concentrations are consistent between the RBS and the XRD measurements.

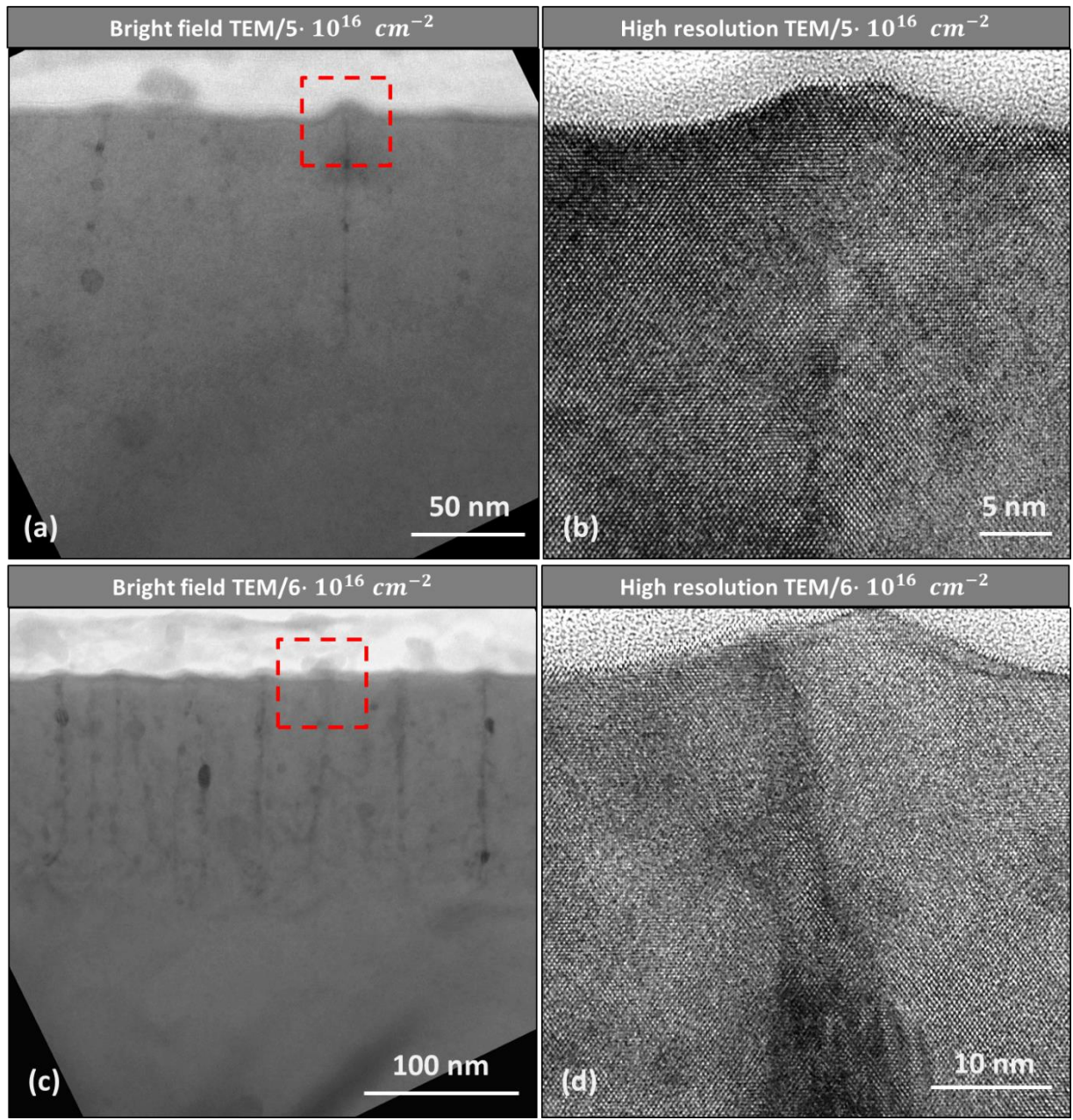

Fig. 4: Cross-section TEM micrographs of the $5 \cdot 10^{16}$ $cm^{-2}$ sample (a,b) and the $6 \cdot 10^{16}$ $cm^{-2}$ sample (c,d).

Cross sectional TEM (XTEM) micrographs for the $5 \cdot 10^{16}$ $cm^{-2}$ and $6 \cdot 10^{16}$ $cm^{-2}$ samples are shown in Fig. 4. In Fig. 4(a) of the $5 \cdot 10^{16}$ $cm^{-2}$ sample, the crystal has high quality with occasional defects. The high resolution XTEM figure of this sample (Fig. 4(b)) confirms the highly ordered Ge-Sn lattice arrangement of the crystal columns. The defects are in the form of thin vertical threads extending from a depth of ∼110 nm to the sample surface. At the end of the threads on the surface are small bumps that appear to be an extrusion of the alloy out from the surface. At the higher implant dose of $6 \cdot 10^{16}$ $cm^{-2}$, Fig. 4(c) shows an increase in the defect density occurs as well as in the diameter of the 'defect threads' which can be seen clearly in the high resolution image in Fig. 4(d).

The origin of these defects is unclear. The vertical orientation of such defects and the possible extrusion or precipitation of material have similarities to the conventional cellular breakdown phenomenon during rapid solidification of a molten layer [26,27]. However, we do not believe this is the case for the following reasons. Such cellular breakdown behaviour is well documented in the literature of laser annealing of semiconductors [28] and occurs when the impurity concentration is typically orders of magnitude above the equilibrium solubility. It results from instability in the melt front under conditions when there is considerable segregation of impurity at the moving melt-solid interface. Excessive segregation causes lateral perturbations in impurity content at the interface and hence lowering of the melting point at regions of high impurity content in the melt. The final outcome is breakdown of the melt front and columns rich in impurity and heavily defective [26]. However, our Ge-Sn system does not exhibit any significant segregation of Sn at the melt-solid interface nor any subsequent surface segregation in any dose or energy regime following PLM. Because such behaviour is expected to be a precursor to cellular breakdown, we do not believe that cellular breakdown is the origin of the defect lines in Fig. 4.

What then is the origin of the threading defects? We believe that such defects are related to the relaxation of the Ge-Sn layer. They are not conventional misfit dislocations that are associated with strain relaxation at elevated temperatures [29,30] since the non-equilibrium PLM process does not give enough time for them to develop at elevated temperature. The Ge-Sn layer leads to relaxation and necessary defect generation to accommodate the large Ge-Sn lattice parameters for growth on a Ge substrate. If this is correct then there may be a level of critical integrated stress in the solidifying Ge-Sn material that, when exceeded, leads to relaxation and defect formation even in highly non-equilibrium PLM. It is likely that such relaxation occurs close to room temperature following completion of the ultra-rapid PLM process but how this might occur is unknown.

Finally, normalized PL curves are shown in Fig. 5. The pristine Ge exhibits emissions at the expected wavelength of 1450 nm and 1880 nm, which correspond to the direct and indirect band gaps, respectively. It should be noted that the spectrum from pure Ge is dominated by the indirect band gap emission due to the high quality of the bulk material, but epitaxial Ge and GeSn films typically exhibit PL spectra dominated by emission from the direct band gap [31,32]. All of the 350 keV GeSn samples show longer emission wavelength compared to the pristine Ge, which we attribute to the incorporation of Sn into the Ge lattice. In the case of the GeSn samples, the spectra show shoulders at the peak wavelengths from pristine Ge, which is due to PL response from the underlying Ge below the depth of the implantation.

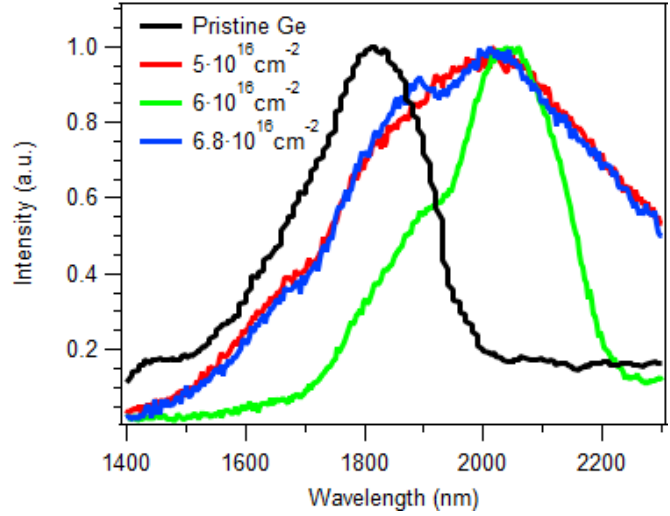

Fig. 5: Photoluminescence of samples implanted at the energy of 350 keV.

In order to interpret the emission spectra from the GeSn samples, it is helpful to first look at the expected band gap values. The compositional dependence of the band gaps at the Γ and L points in GeSn alloys has been studied extensively [33-38]. Generally, the band gaps can be calculated using the expression

$$E_i(x) = xE_i^{Sn} + (1-x)E_i^{Ge} - b_i x(1-x), \qquad \text{(Eq. 1)}$$

where $E_i$ is the band gap at the $i$ point (in this case Γ or L), $E_i^{Sn}$ is the band gap of α-Sn at that point, $E_i^{Ge}$ is the band gap of Ge at that point, $b_i$ is the bowing parameter for that point, and $x$ is the Sn concentration in the alloy. The bowing parameters for the two gaps in question have been calculated from fits to data obtained from photoluminescence measurements by a number of groups, but there is still some disagreement on the values. This is mainly due to the fact that the effects of strain and composition on the band gap cannot be separated, and the measurements were nearly all performed on samples with some residual strain. Additionally, there is some evidence that the bowing parameter may be compositionally-dependent, leading to a more complicated expression than Eq. 1 [38]. Nevertheless, we can choose a bowing parameter from the literature that was obtained from samples with similar Sn concentrations and strains. In this case, we use the bowing parameters obtained from Ref. [37]. The calculated gap energies and their corresponding wavelengths are presented in Table I.

| Implantation dose (cm$^{-2}$) | Peak Sn concentration (%) | $E_\Gamma$ (eV) | $\lambda_\Gamma$ (nm) |
|---|---|---|---|
| $5 \cdot 10^{16}$ | 4.8 | 0.62 | 2000 |
| $6 \cdot 10^{16}$ | 5.25 | 0.61 | 2030 |
| $6.8 \cdot 10^{16}$ | 6 | 0.58 | 2140 |

Table I. List of band gaps and wavelengths at the Γ- and L- points in the samples as calculated using Eq. 1 with the bowing parameters $b_\Gamma = 2.46$ eV and $b_L = 1.03$ eV taken from Ref. [37].

The PL spectrum obtained from the $5 \cdot 10^{16}$ cm$^{-2}$ sample has a broad peak that is centered on 2015 nm. According to Eq. 1 the Sn concentration equivalent to this wavelength is 5.1 at. %. To the left of the main peak, there is still appreciable PL signal beyond the wavelength corresponding to the indirect band gap in pure Ge. It is important to note that the Sn concentration is not constant throughout the GeSn layer, so it is likely that there is some PL signal from regions where there is crystalline GeSn with Sn concentrations lower than the peak, which produces emission at a shorter wavelength. Emission at shorter wavelengths could also be due to splitting of the valence band due to residual strain in the material [39]. There is also some weak emission to the right of the peak, which we attribute to emission from the indirect band gap.

In the case of the $6.8 \cdot 10^{16}$ $cm^{-2}$ sample, there are two distinct emission peaks at 1890 $nm$ and 2010 $nm$, which are below the calculated wavelengths shown in Table I. We attribute this to the formation of defects in the GeSn layer during the PLM process. At this higher implant dose, both the volume and the density of the threading defects observed in Fig. 4 increase. The observed PL emission most likely comes from regions with lower Sn content but high crystal quality, while the PL from the more defective, higher Sn content region is weak due to an increase in non-radiative Shockley-Read-Hall (SRH) recombination due to defects. The Sn content in this sample is very near the predicted indirect-to-direct band gap transition, but there was no large change in PL intensity that would be expected from the shift to direct band gap. The peak at 2010 $nm$ is thus likely to be due to direct band gap emission from the regions of sufficient quality, and the wavelength corresponds to a Sn content of 5 at%. It should be noted that the peak appears in the same position as the peak in the spectrum obtained from the $5 \cdot 10^{16}$ cm$^{-2}$ sample.

The PL spectrum from the $6 \cdot 10^{16}$ cm$^{-2}$ sample has peaks at 1900 nm and 2045 nm. The main emission peak is centered on 2045 nm, suggesting that there is a high-quality region with Sn content ~6 at. %, comparable to the values obtained by RBS and XRD. The peak at 1900 $nm$ is attributed to the splitting of the valence band due to strain. The width of the peaks obtained from this sample is much narrower than those from the other GeSn samples. The laser pump power, the slit width, and the detector gain were all held constant for these experiments, so there should be no difference due to the experimental setup. Thus, this difference in width must be due to the material itself. In this case, we likely have a better-defined region with constant Sn content that gives rise to the PL signal. This could be due to differences in crystal formation during the PLM process, although it is not clear exactly how. However, it does suggest that it is possible to form highly-crystalline regions with Sn content up to ~6 at.% and that, past that point, relaxation of the lattice through defect generation has a detrimental effect on crystal quality. A possible solution to improve crystal quality of the samples is to re-amorphise partly the relaxed-GeSn layer by a second implantation of Sn. A subsequent PLM of this layer to melt just the amorphous layer to re-crystallise it from the underlying relaxed Ge-Sn seed may cause it to regrow in a near defect-free manner.

## Conclusion

Ion implantation followed by pulsed laser melting is shown to produce a Ge-Sn alloy with the substitutional Sn content of ~6 at.% and a high degree of strain relaxation. All the samples have photo-luminescence responses at wavelengths longer than that of pure Ge, indicating the active incorporation of Sn in the lattice. The XTEM shows that the non-equilibrium threading defects are the mechanism for the efficient relaxation of the Ge-Sn layer. These defects are neither the threading dislocations found in heterogeneous epitaxy of material nor the conventional cellular breakdown related defects in a PLM process. These threading-type defects appear to deteriorate the photon emission, in particular in the $6.8 \cdot 10^{16}\ cm^{-2}$ sample. A possible solution to remove such defects is to re-amorphise and PLM part of the Ge-Sn alloy so that the melted layer can regrow from the fully-relaxed Ge-Sn seed. Such a process would be able to produce a relaxed Ge-Sn alloy with a quality suitable for device applications.

## Acknowledgement

The authors would like to acknowledge the Australian Research Council for the funding support, the National Collaborative Research Infrastructure Strategy for the access to the Australian National Fabrication Facility and the Heavy Ion Accelerator Facility. J. Mathews would also like to acknowledge funding support from an Air Force Office of Scientific Research Young Investigator Award, Grant number FA9550-17-1-0146.